\documentclass[%
 amsmath,amssymb,
 preprint,
 superscriptaddress,
]{revtex4-1}

\usepackage{graphicx}% Include figure files
\usepackage{dcolumn}% Align table columns on decimal point
\usepackage{bm}% bold math
\usepackage[utf8]{inputenc}
\usepackage[T1]{fontenc}
\usepackage{mathptmx}
\usepackage{etoolbox}
\usepackage{url}            % simple URL typesetting
\usepackage{subfiles}
\usepackage{array}
\newcolumntype{P}[1]{>{\centering\arraybackslash}m{#1}}
\usepackage{microtype}
\usepackage{amsmath}
\usepackage{amssymb}
\usepackage{mathtools}
\usepackage{txfonts}
\setcitestyle{super}

\begin{document}

\title{Space-time structured plasma waves}% Force line breaks with \\

\author{J.P. Palastro}
\email{jpal@lle.rochester.edu}
\affiliation{
University of Rochester, Laboratory for Laser Energetics, Rochester, New York 14623-1299 USA}
\author{K.G. Miller}
\affiliation{
University of Rochester, Laboratory for Laser Energetics, Rochester, New York 14623-1299 USA}
\author{R.K. Follett}
\affiliation{
University of Rochester, Laboratory for Laser Energetics, Rochester, New York 14623-1299 USA}
\author{D. Ramsey}
\affiliation{
University of Rochester, Laboratory for Laser Energetics, Rochester, New York 14623-1299 USA}
\author{K. Weichman}
\affiliation{
University of Rochester, Laboratory for Laser Energetics, Rochester, New York 14623-1299 USA}
\author{A.V. Arefiev}
\affiliation{
Department of Mechanical and Aerospace Engineering, University of California at San Diego, La Jolla, California 92093, USA}
\author{D.H. Froula}
\affiliation{
University of Rochester, Laboratory for Laser Energetics, Rochester, New York 14623-1299 USA}

\date{\today}
\begin{abstract}
Electrostatic waves play a critical role in nearly every branch of plasma physics from fusion to advanced accelerators, to astro, solar, and ionospheric physics. The properties of planar electrostatic waves are fully determined by the plasma conditions, such as density, temperature, ionization state, or details of the distribution functions. Here we demonstrate that electrostatic wave packets structured with space-time correlations can have properties that are independent of the plasma conditions. For instance, an appropriately structured electrostatic wave packet can travel at any group velocity, even backward with respect to its phase fronts, while maintaining a localized energy density. These linear, propagation-invariant wave packets can be constructed with or without orbital angular momentum by superposing natural modes of the plasma and can be ponderomotively excited by space-time structured laser pulses like the flying focus.

\end{abstract}

                      
\maketitle

A defining characteristic of plasma is its ability to exhibit collective motion. This motion often manifests as coordinated oscillations of the constituent particles, mediated by their mutual electrostatic attraction or repulsion. The oscillations, or electrostatic waves, play a critical role in nearly every branch of plasma physics. In fusion, electrostatic waves can be both a feature, providing a means to measure plasma conditions \cite{Behn89,Glenzer1999,Froula02,Bindslev06,Korsholm11,Hansen21}, and an impediment, growing unstably to the point of disrupting plasma confinement and heating \cite{Kruer96,Porkolab77,Tang78,Jenko01,Michel12,Proll12,Michel15,Montgomery2016}. Advanced accelerators harness electrostatic waves to accelerate electrons to relativistic energies over short distances, with the ultimate goal of miniaturizing radiation sources and particle colliders \cite{Tajima1979,Lu2007,Esarey2009,Gonsalves2019,Miao2022,Caldwell2009,Assmann2014,Litos2014,Gessner2016,Joshi2018}. As a final, naturally occurring example, the mode conversion of electrostatic waves driven by fast electrons can explain the emission of type III radio bursts from the solar wind \cite{Lin1986,Ergun1998,Reid2014}.

In each of these systems, the evolution of electrostatic waves impacts performance, dynamics, or observations. The evolution of planar electrostatic waves, i.e., waves having a single frequency $\omega$ and wavevector $\mathbf{k}$, is fully determined by the plasma conditions through the dispersion relation $\varepsilon(\omega,\mathbf{k}) = 0$. More specifically, the phase velocity $\mathbf{v}_p = [\omega(\mathbf{k})/k]\mathbf{e}_\mathbf{k}$ can depend on the density, temperature, ionization states, or details of the distribution functions. Physically occurring electrostatic waves exist as superpositions of plane waves with amplitudes and phases imposed by a driver, such as an intense laser pulse or charged particle beam. A typical driver excites the wave packets without introducing correlations in $(\omega,\mathbf{k})$ space. As a result, the wave packets retain properties similar to those of a plane wave. However, electrostatic wave packets can also be driven so that they feature correlations in $(\omega,\mathbf{k})$ space. With appropriate structuring, these correlations can produce emergent properties that are independent of the plasma conditions.

The structuring of {\it electromagnetic} waves with space-time correlations has provided new opportunities for laser-based applications and basic science \cite{Howard19,Debus2019,Palastro2020,Caizergues2020,DiPiazza2021,Ramsey2022,Kabacinski2023,Simpson2023}. This has motivated the development of optical techniques for creating structured light, such as propagation-invariant \cite{Longhi03, Kondakci2017,Kondakci2019,Li2020,Li2020a,Yessenov2022,Besieris22}, flying focus \cite{Sainte-Marie2017,Froula2018,Palastro2020,Jolly2020,Simpson2022,Ambat2023}, and arbitrarily-structured-laser (ASTRL) pulses \cite{Pierce2023}. While these techniques cannot be directly applied to electrostatic waves, much of the mathematical formalism carries over: at a fundamental level, all waves evolve according to a wave equation. Thus, by using an appropriate driver, one can construct electrostatic analogs to propagation-invariant, flying focus, or ASTRL pulses. 

This manuscript introduces the concept of space-time structured plasma waves. A space-time structured plasma wave (STP) can be constructed, with our without orbital angular momentum, by superposing natural electrostatic modes of a plasma with a particular correlation in $(\omega,\mathbf{k})$ space. As an example, we focus on the special case of a linear, propagation-invariant electrostatic wave packet with a group velocity that is independent of the plasma conditions. The excitation of such an STP can be achieved experimentally by using the ponderomotive force of a structured laser pulse like a flying focus. STPs offer a new class of collective excitations that may provide additional control over dynamics such as wave-particle interactions, particularly in situations where the driver can be structured.

\begin{figure}
\includegraphics[width=0.75\linewidth]{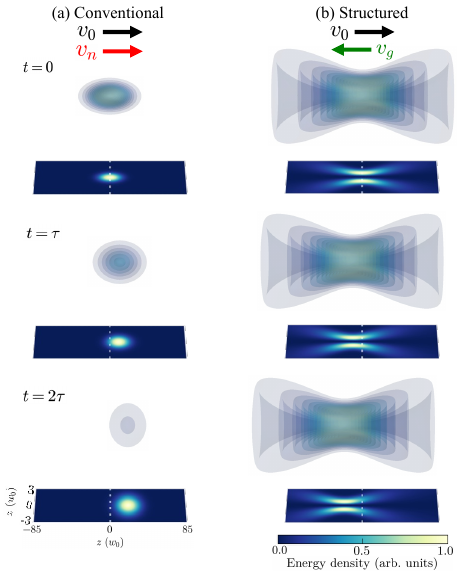}
\caption{Evolution of the cycle-averaged energy density $\varepsilon_0\langle k_0^2\phi^2\rangle$ for a conventional and space-time structured plasma wave (STP). The conventional plasma wave (left) diffracts as it propagates from left to right at a nominal group velocity $v_n$ determined by the plasma conditions. The peak energy density of the STP (right) travels in the opposite direction as the nominal group velocity and phase velocity while maintaining a constant spatiotemporal profile. In both cases, $k_0w_0 = 20$. The STP has $v_g = -v_n$ and $\ell = 1$. For the conventional plasma wave, $Z_0 = 4.5w_0$ and $\ell = 0$. Space is normalized by $w_0$ and time by $\tau = \omega_0w_0^2/2u^2$ [see Table I and Eqs. \eqref{eq:waveeqPhiST} and \eqref{eq:waveeqPhiCON}]. The contours have the same normalization, while each projection is normalized to its maximum.}
\label{fig:f1}
\end{figure}

Figure \ref{fig:f1} contrasts a conventional, localized plasma wave with an STP. The conventional plasma wave propagates with a group velocity determined by the plasma conditions. As the wave propagates, diffraction causes a rapid drop in the peak energy density. The STP travels at a velocity that is independent of the plasma conditions and maintains its profile, and peak energy density, over an extended distance. In this example, the peak energy density travels in the opposite direction as the phase fronts and the nominal group velocity.

The formulation of STPs will be presented for pure electrostatic waves in the absence of external fields. Pure electrostatic plane waves have a wavevector that is parallel to their electric field $\mathbf{E}$ and have no magnetic field, i.e., $\mathbf{k} \times \mathbf{E} = 0$. These waves are completely described by their electrostatic potential. The electrostatic potential $\phi$ of a plasma wave packet can be expressed as a superposition of plane waves constrained by the dispersion relation:
\begin{equation}\label{eq:phi}
\phi(\mathbf{x},t) = \int \phi_0(\omega,\mathbf{k})e^{i(\mathbf{k}\cdot\mathbf{x}-\omega t)} \delta[\varepsilon(\omega,\mathbf{k})]d\mathbf{k}d\omega,
\end{equation}
where $\delta$ is the Dirac delta function and the conditions $\phi_0(\omega,\mathbf{k}) = \phi_0^*(-\omega,-\mathbf{k})$ and $\varepsilon(\omega,\mathbf{k}) = \varepsilon^*(-\omega,-\mathbf{k})$  ensure that $\phi$ is real. The constraint imposed by the dispersion relation collapses one of the integrals in Eq. \eqref{eq:phi} and is typically used to write the frequency in terms of the wavevector, i.e., $\omega = \omega(\mathbf{k})$ with  $\varepsilon[\omega(\mathbf{k}),\mathbf{k}] = 0$ implied.

Aside from the dispersion relation, an additional constraint $C(\omega,\mathbf{k})$ can be applied by writing
\begin{equation}\label{eq:newphi}
\phi_0(\omega,\mathbf{k})=\bar{\phi}_0(\omega,\mathbf{k})\delta[C(\omega,\mathbf{k})].
\end{equation}
The most general form of an STP uses $C(\omega,\mathbf{k})$ to introduce correlations in $(\omega,\mathbf{k})$ space. Motivated by propagation invariant and flying focus laser pulses \cite{Longhi03,Kondakci2017,Ramsey2022}, the constraint is chosen here to allow for an arbitrary, specified group velocity $v_g$: 
\begin{equation}\label{eq:C}
C(\omega,\mathbf{k})= \frac{v_g}{(\omega_0 - v_gk_0)}[(\omega-v_gk_z)^2 - (\omega_0-v_gk_0)^2].
\end{equation}
Upon setting Eq. \eqref{eq:C} equal to zero in accordance with the delta function, one can verify that 
\begin{equation}\label{eq:vg}
\frac{\partial \omega}{\partial k_z} = v_g.
\end{equation}
Substituting Eq. \eqref{eq:newphi} into Eq. \eqref{eq:phi} and applying the constraint provides the electrostatic potential of the STP:
\begin{equation}\label{eq:stpphi}
\phi(\mathbf{x}_{\perp},\eta,\xi) = \tfrac{1}{2}e^{ik_0\eta}\Phi(\mathbf{x}_{\perp},\xi) + \text{c.c.},
\end{equation}
where $\eta = z - v_0 t$, $v_0 = \omega_0/k_0$, $\xi = z - v_g t$, 
\begin{equation}\label{eq:bigphi}
\Phi(\mathbf{x}_{\perp},\xi) = \int \bar{\Phi}(\Omega,\mathbf{k}_{\perp})e^{i\mathbf{k}_{\perp}\cdot\mathbf{x}_{\perp} + i\Omega\xi/v_g}\delta(\varepsilon)d\mathbf{k_{\perp}}d\Omega,
\end{equation}
$\Omega = \omega - \omega_0$, and $\bar{\Phi}(\Omega,\mathbf{k_{\perp})} = \bar{\phi}_0(\omega_0+\Omega,\mathbf{k}_{\perp},k_0 + \Omega/v_g)$. Equation \eqref{eq:stpphi} demonstrates that an electrostatic potential constructed with the correlation $C(\omega,\mathbf{k})$ has phase fronts that travel at the velocity $v_0$ and an envelope $\Phi$ that travels at the group velocity $v_g$.

Thus far, the formulation has been relatively abstract. To make the concept more tangible, examples will be presented for a non-relativistic, non-flowing plasma composed of electrons and a single ion species. The dispersion relation for such a plasma can be derived using the Vlasov-Poisson system of equations and is given by
\begin{equation}\label{eq:eps}
\varepsilon(\omega,\mathbf{k}) = 1 + \chi_e(\omega,\mathbf{k}) + \chi_i(\omega,\mathbf{k}) = 0,
\end{equation}
where
\begin{equation}
\chi_s(\omega,\mathbf{k}) = \frac{\omega_{ps}^2}{k^{2}}\int \frac{\mathbf{k}\cdot \nabla_{\mathbf{v}} f_{s}}{\omega-\mathbf{k}\cdot\mathbf{v}}d\mathbf{v}
\end{equation}
is the susceptibility for species $s$, $\omega_{ps} = (q_s^2n_s/\varepsilon_0m_s)^{1/2}$ is the plasma frequency, $n_s$ is the density, and $f_s=f_s(\mathbf{v})$ is the velocity distribution function. Equation \eqref{eq:eps} predicts the existence of two elementary plasma waves: a high-frequency electron plasma wave and a low-frequency ion-acoustic wave. 

The dispersion relation for electron plasma waves can be found in the limit that the phase velocity is much greater than the electron thermal velocity, i.e., $v_p \gg v_{Te}$, where $v_{Ts} = [\int v_k^2f_sd\mathbf{v}]^{1/2}$ and $v_k = \mathbf{e}_\mathbf{k}\cdot\mathbf{v}$. In this limit, Eq. \eqref{eq:eps} reduces to
\begin{equation}\label{eq:epw}
\varepsilon(\omega,\mathbf{k}) \approx 1 - \frac{\omega_{pe}^2}{\omega^2} - 3k^2\lambda_{De}^2
\end{equation}
where $\lambda_{De} = v_{Te}/\omega_{pe}$ is the electron Debye length. The dispersion relation for ion-acoustic waves can be found in the opposite limit where the phase velocity is much smaller than the electron thermal velocity, i.e., $v_p \ll v_{Te}$. Here, Eq. \eqref{eq:eps} reduces to
\begin{equation}\label{eq:iaw}
\varepsilon(\omega,\mathbf{k}) \approx 1 + \frac{1}{k^2\lambda_{De}^2} -  \frac{\omega_{pi}^2}{\omega^2}.
\end{equation}
In both cases, Landau damping has been neglected. This is a good approximation when $k\lambda_{De} \lesssim 0.2$ or $v_p/v_{Ti} \gg 1$ for electron plasma and ion-acoustic waves, respectively.

Without the constraint $C(\omega,\mathbf{k})$, the plasma waves travel at a group velocity determined by the plasma conditions. Solving for the frequency in Eqs. \eqref{eq:epw} and \eqref{eq:iaw} yields $\omega(k) = (\varpi^2 + u^2k^2)^{1/2}$ and the group velocity
\begin{equation}
\frac{\partial \omega}{\partial k_z} = \frac{u^2k_z}{\omega}, 
\end{equation}
where the values of $\varpi$ and $u$ for each wave are defined in Table I. With the constraint $C(\omega,\mathbf{k})$ [Eq. \eqref{eq:C}], the relation $\omega(k) = (\varpi^2 + u^2k^2)^{1/2}$ still holds, but now the transverse wavenumber is a function of the longitudinal wavenumber (or frequency), such that
\begin{equation}
\frac{\partial \omega}{\partial k_z} = \frac{u^2}{\omega}\left(k_z + \frac{1}{2}\frac{\partial k_{\perp}^2}{\partial k_z}\right) = v_g,
\end{equation}
where $k_z = k_0 + \Omega/v_g$ has been used. Thus the velocity $v_g$ is completely independent of the plasma conditions.

\begin{table}[b]
\caption{\label{tab:table1}%
Variable definitions used for the electron plasma wave (EPW) and ion-acoustic wave (IAW). $c_s= \omega_{pi}\lambda_{De}$ is the sound speed.
}
\begin{ruledtabular}
\begin{tabular}{lcc}
\textrm{Variable}&
\textrm{EPW} &
\textrm{IAW} \\
\hline
$\varpi$ & $\omega_{pe}$ & 0 \\
$u$ & $\sqrt3v_{Te}$ & $c_s$ \\
\end{tabular}
\end{ruledtabular}
\end{table}

Using the explicit expressions for $\varepsilon$ from Eqs. \eqref{eq:epw} and \eqref{eq:iaw}, the delta function enforcing the dispersion relation can be written in the general form
\begin{equation}\label{eq:epscon}
\delta(\varepsilon) = \left| \frac{\partial \Omega}{\partial\varepsilon}  \right|_{\Omega = \Omega_n} \delta[\Omega - \Omega_n(k_{\perp})],
\end{equation}
where 
\begin{equation}
\begin{aligned}\label{eq:Omegan}
\Omega_n&(k_{\perp}) = -k_0v_gv_0\left(\frac{v_g-v_n}{v_g^2-u^2} \right) \\ &+ \left[ (k_0v_gv_0)^2 \left(\frac{v_g-v_n}{v_g^2-u^2} \right)^2 +  \left(\frac{v_g^2u^2k_{\perp}^2}{v_g^2-u^2}\right)\right]^{1/2}
\end{aligned}
\end{equation}
and $v_n = u^2/v_0$. In arriving at Eq. \eqref{eq:Omegan}, the choice was made to set $\omega_0 = (\varpi^2 + u^2k_0^2)^{1/2}$. With this choice, $v_0 = \omega_0/k_0$ equals the phase velocity of the plasma wave in the plane-wave limit $v_g \rightarrow 0$, and $v_n$ equals the nominal group velocity in the absence of space-time structuring. Applying Eq. \eqref{eq:epscon} in Eq. \eqref{eq:bigphi} collapses the integral over $\Omega$, leaving only the integral over $\mathbf{k}_{\perp}$:
\begin{equation}\label{eq:bigphired}
\Phi(\mathbf{x}_{\perp},\xi) = \int \tilde{\Phi}(\mathbf{k}_{\perp})e^{i\mathbf{k_{\perp}}\cdot\mathbf{x}_{\perp} + i\Omega_n(k_{\perp})\xi/v_g}d\mathbf{k_{\perp}},
\end{equation}
where $\tilde{\Phi}(\mathbf{k_{\perp}}) = |\partial_{\varepsilon}\Omega|_{\Omega=\Omega_n} \bar{\Phi}[\Omega_n(k_{\perp}),\mathbf{k}_{\perp}]$. The function $\tilde{\Phi}(\mathbf{k}_{\perp})$ determines the spatiotemporal profile of the arbitrary group velocity plasma wave.

Analytic expressions for the spatiotemporal profile can be found in the ``paraxial'' approximation, i.e., when the condition
\begin{equation}\label{eq:parax}
k_{\perp}^2 \ll \frac{\:v_0(v_g-v_n)^2}{v_n(v_g^2-u^2)}k_0^2
\end{equation}
is satisfied. Upon using this condition, $\Omega_n$ simplifies to
\begin{equation}\label{eq:Omeganpar}
\Omega_n(k_{\perp}) \approx 
\left(\frac{v_gv_n}{v_g-v_n}\right)\frac{k_{\perp}^2}{2k_0}.
\end{equation}
With the quadratic dependence of $\Omega_n$ on $k_{\perp}$, a natural choice for 
$\tilde{\Phi}(\mathbf{k}_{\perp})$ is a superposition of Laguerre-Gaussian modes, i.e., 
\begin{equation}\label{eq:Phitilde}
\tilde{\Phi}(\mathbf{k}_{\perp}) = \sum_{p,\ell} \tilde{\Phi}_{p\ell}\kappa^{\ell}L_p^{\ell}(\kappa)\mathrm{exp}(-\tfrac{1}{2}\kappa^2)e^{i\ell\theta_k},
\end{equation}
where $\kappa = k_{\perp}w_0/\sqrt{2}$, $w_0$ characterizes the transverse width, $L_p^{\ell}$ is a generalized Laguerre polynomial, and $\theta_k$ is the azimuth in transverse wavenumber space. The spatiotemporal profile of the STP is then given by
\begin{equation}\label{eq:AnSol}
\begin{aligned}
\Phi(&\mathbf{x}_{\perp},\xi) = \sum_{p,\ell} \Phi_{p\ell} \frac{w_0}{w}\Bigl( \frac{\sqrt{2} r}{w}\Bigr)^\ell L^\ell_p\Bigl(\frac{2r^2}{w^2}\Bigr) 
e^{i\ell\theta}
\\
&\text{exp}\left[ - \Bigl(1-i\frac{\xi}{\xi_0}\Bigr)\frac{r^2}{w^2} -i(2p+\ell+1) \text{arctan}\frac{\xi}{\xi_0}\right],
\end{aligned}
\end{equation}
where $w(\xi) = w_0[1+(\xi/\xi_0)^2]^{1/2}$,
\begin{equation}\label{eq:xi0}
\xi_0 = \frac{(v_g-v_n)k_0w_0^2}{2v_n},
\end{equation}
$r=(x^2+y^2)^{1/2}$, $\theta = \mathrm{arctan}(y/x)$ is the azimuth in configuration space, and constant factors have been absorbed into the amplitudes $\Phi_{p\ell}$. The profile of the STP advects at the group velocity $v_g$, has a characteristic duration $\xi_0/v_g$, and can have any orbital angular momentum value $\ell$.

The analysis so far has demonstrated that an arbitrary group velocity STP can be constructed theoretically, but has not provided a prescription for how to do so in practice. Plasma waves can either exist as thermal fluctuations or be driven by external forces. Thermal fluctuations have no correlations in $(\omega,\mathbf{k})$ space, and other than having to satisfy $\varepsilon(\omega,\mathbf{k})=0$, $k_{\perp}$ and $k_z$ are completely independent, i.e., $\partial k_{\perp} / \partial k_z = 0$. As a result, an STP must be driven by external forces, such as those exerted by particle beams or electromagnetic waves. In the presence of an external force $\mathbf{F}(\mathbf{x},t)$, the potential of a generic electrostatic wave is given by
\begin{equation}\label{eq:phiF}
\phi(\mathbf{x},t) = \int \frac{i}{ek}\frac{\chi_e(\omega,\mathbf{k})}{\varepsilon(\omega,\mathbf{k})} [\mathbf{e}_k \cdot \hat{\mathbf{F}}(\omega,\mathbf{k})]e^{i(\mathbf{k}\cdot\mathbf{x}-\omega t)} d\mathbf{k}d\omega,
\end{equation}
where $e$ is the elementary charge. Resonant excitation of an STP requires that the force $\mathbf{F}(\mathbf{x},t)$ be a function of space and time in the combinations $\eta = z - v_0 t$ and $\xi = z - v_g t$. 

%In order to resonantly drive an STP, the force $\mathbf{F}(\mathbf{x},t)$ must be a function of space and time in the combinations $\eta = z - v_0 t$ and $\xi = z - v_g t$. 

Electromagnetic waves provide a flexible option for driving STPs. Laser pulses, in particular, can exhibit correlations between two or more degrees of freedom, including polarization, orbital angular momentum, and spatio-spectral content, and can interact in geometries ranging from co- to counter-propagating. When the frequencies of the electromagnetic waves are much greater than $\omega_0$, the disparity of time scales allows for a cycle-averaging over their periods. The end result is a ``ponderomotive guiding center'' equation of motion with the effective force
\begin{equation}\label{eq:pond}
\mathbf{F}(\mathbf{x},t) = -\frac{1}{2}m_e c^2\nabla \langle \bold{a} \cdot \bold{a} \rangle
\end{equation}
where $\bold{a}(\mathbf{x},t)  = e\bold{A}(\mathbf{x},t) /m_ec$ is the total normalized vector potential of the electromagnetic waves, satisfying $|\bold{a}|\ll 1 $, and $\langle \rangle$ represents a cycle-average.

An STP can be resonantly excited using a superposition of two flying focus pulses. Flying focus pulses feature an intensity peak that can travel at any velocity $v_f$, while maintaining a near-constant spatiotemporal profile. The interference of two flying focus pulses with $v_f = v_g$ and distinct frequencies and wavenumbers satisfying $\omega_1 - \omega_2 = \omega_0$ and $\mathbf{e}_z \cdot (\mathbf{k}_{1} - \mathbf{k}_{2}) = k_0$ produces the ponderomotive force neccessary to resonantly drive an STP. Specifically, the superposition
\begin{equation}\label{eq:affs}
\bold{a}(\mathbf{x},t) = \tfrac{1}{2} \sum_{j\in(1,2)} \bm{a}_j(\mathbf{x}_{\perp},\xi)e^{i(k_{j}z - \omega_j t)} + \mathrm{c.c.},
\end{equation}
where $\bm{a}_j(\mathbf{x}_{\perp},\xi)$ is the envelope of each pulse, results in a ponderomotive force term
\begin{equation}\label{eq:pondrelevant}
\mathbf{F}_d(\mathbf{x}_{\perp},\eta,\xi) = -\frac{i}{8}m_e c^2 k_0 (\bm{a}_1 \cdot \bm{a}_2^{*})e^{ik_0\eta}\mathbf{e}_z + \mathrm{c.c.}. 
\end{equation}
In writing Eq. \eqref{eq:affs}, it has been assumed that the durations of the flying focus pulses are much longer than their periods $2\pi/\omega_j$.

Without further specification of the $\bm{a}_j$, the electrostatic potential of the driven STP is given by $\phi(\mathbf{x}_{\perp},\eta,\xi) = \tfrac{1}{2}e^{ik_0\eta}\Phi(\mathbf{x}_{\perp},\xi) + \text{c.c.},$ with
\begin{equation}\label{eq:bigphidriven}
\Phi(\mathbf{x}_{\perp},\xi) = \int S(\Omega,\mathbf{k}_{\perp})e^{i\mathbf{k}_{\perp}\cdot\mathbf{x}_{\perp} + i\Omega\xi/v_g}d\mathbf{k_{\perp}}d\Omega
\end{equation}
and
\begin{equation}\label{eq:bigess}
\begin{split}
S(\Omega,\mathbf{k}_{\perp}) &= \frac{m_ec^2}{32\pi^3e}\frac{\mathrm{s}(v_0-v_g)}{|v_g|}\frac{k_0\chi_e(\omega,\mathbf{k})}{k\varepsilon(\omega,\mathbf{k})} \\ & \qquad \int (\bm{a}_1 \cdot \bm{a}_2^{*})e^{-i\mathbf{k}_{\perp}\cdot\mathbf{x}_{\perp} - i\Omega\xi/v_g}d\mathbf{x_{\perp}}d\xi,
\end{split}
\end{equation}
where $\mathrm{s}$ is the sign function, and $\chi_e$, $\varepsilon$, and $k$, are evaluated at $\omega = \Omega + \omega_0$ and $k_z = k_0 + \Omega/v_g$. Thus, the ponderomotive force of the two flying focus pulses drives an electrostatic potential with phase fronts that travel at $v_0$ and an envelope $\Phi(\mathbf{x}_{\perp},\xi)$ that travels at $v_g$. Note that while the frequency and wavenumber matching conditions, i.e., $\omega_1 - \omega_2 = \omega_0$ and $\mathbf{e}_z \cdot (\mathbf{k}_{1} - \mathbf{k}_{2}) = k_0$, are identical to those required for stimulated Raman or Brillouin scattering (electron and ion-acoustic waves, respectively), excitation of an STP \textit{does not} require instability.

Equations \eqref{eq:pondrelevant} and \eqref{eq:bigess} provide an exact, linear solution for a driven STP in the spectral domain. While these solutions demonstrate the salient physics, they are ``monochromatic,'' that is, they oscillate in $\eta$ with a single period $2\pi/k_0$. More generally, the potential will be a superposition of these solutions, such that
\begin{equation}\label{eq:stpphiBW}
\phi(\mathbf{x}_{\perp},\eta,\xi) = \tfrac{1}{2}e^{ik_0\eta} \int \breve{\Phi}(\mathbf{x}_{\perp},k',\xi)e^{ik'\eta} dk' + \text{c.c.},
\end{equation}
where $k'$ represents a wavenumber shift about the central wavenumber $k_0$ and the envelope of the potential $\Phi(\mathbf{x}_{\perp},\eta,\xi) = \int \breve{\Phi}(\mathbf{x}_{\perp},k',\xi)e^{ik'\eta} dk'$ now depends on $\eta$.

Direct evaluation of Eq. \eqref{eq:phiF} can be challenging. As an alternative, when $\omega_0$ is close to the natural mode frequency of the plasma wave, Eq. \eqref{eq:phiF} can be recast as the configuration-space wave equation
\begin{equation}\label{eq:waveeq}
\left(\partial^2_t +\varpi^2 - u^2 \nabla^2 \right) \phi(\mathbf{x},t) = \pm \frac{1}{8} \omega_0^2(\bm{a}_1 \cdot \bm{a}_2^{*})e^{ik_0\eta} + \mathrm{c.c.},
\end{equation}
where $\phi$ has been normalized by $m_ec^2/e$, and the top and bottom signs are taken for electron plasma and ion-acoustic waves, respectively. If the $\bm{a}_j$ are independent of $\eta$ (or approximately so), Eq. \eqref{eq:waveeq} reduces to
\begin{equation}\label{eq:waveeqPhiST}
\left[2i\kappa \frac{\partial}{\partial \xi} + \frac{u^2 - v_g^2}{u^2}  \frac{\partial^2}{\partial \xi^2} + \nabla_{\perp}^2\right]\Phi_s(\mathbf{x}_{\perp},\xi) = \mp\frac{\omega_0^2(\bm{a}_1 \cdot \bm{a}_2^{*})}{8u^2},
\end{equation}
where $\kappa=k_0(v_n-v_g)/v_n$ and the subscript $s$ refers to the STP. The homogeneous dispersion relation for Eq. \eqref{eq:waveeq} is given by Eq. \eqref{eq:Omegan} and the homogeneous, paraxial solutions by Eq. \eqref{eq:AnSol}. The evolution of $\Phi_s(\mathbf{x}_{\perp},\xi)$  contrasts that of a conventional plasma wave for which Eq. \eqref{eq:waveeq} is often simplified as 
\begin{equation}\label{eq:waveeqPhiCON}
\left[2i\frac{\omega_0}{u^2}\frac{\partial}{\partial t} +  \frac{\partial^2}{\partial \zeta^2} + \nabla_{\perp}^2\right]\Phi_c(\mathbf{x}_{\perp},\zeta,t) = \mp\frac{\omega_0^2(\bm{a}_1 \cdot \bm{a}_2^{*})}{8u^2},
\end{equation}
where $\zeta = z - v_nt$, the subscript $c$ refers to a conventional plasma wave, and $|\partial_t \Phi_c| \ll |\omega_0 \Phi_c|$ has been assumed. In Fig. \ref{fig:f1} the homogeneous solutions to Eqs. \eqref{eq:waveeqPhiST} and \eqref{eq:waveeqPhiCON} are compared for the initial conditions $\Phi_s(\mathbf{x}_{\perp},0) = \Phi_0 (\sqrt2r/w_0)\exp{(-r^2/w_0^2)e^{i\theta}}$ and $\Phi_c(\mathbf{x}_{\perp},\zeta,0) = \Phi_0 \exp{(-r^2/w_0^2-\zeta^2/Z_0^2)}$, respectively. The $\ell = 1$ mode was chosen for the STP to illustrate its ability to carry orbital angular momentum.

As a final note, the delta function that enforces the dispersion relation in Eq. \eqref{eq:bigphi} can also be written in terms of the perpendicular wavenumber $k_{\perp}$: 
\begin{equation}\label{eq:epscon2}
\delta(\varepsilon) = \left| \frac{\partial k_{\perp}}{\partial\varepsilon}  \right|_{k_{\perp} = k_{\perp,n}} \delta[k_{\perp} -  k_{\perp,n}(\Omega)],
\end{equation}
where 
\begin{equation}
\begin{aligned}\label{eq:kperpn}
k_{\perp,n}(\Omega)& = \frac{1}{u} \left[ \Omega^2\left(1-\frac{u^2}{v_g^2}\right) +2\omega_0\Omega\left(1-\frac{u^2}{v_0v_g}\right)  \right]^{1/2}.
\end{aligned}
\end{equation}
This allows one to write $\tilde{\Phi}$ as a function of $\Omega$ instead of $\mathbf{k}_\perp$ when evaluating $\Phi(\mathbf{x}_\perp,\xi)$ in Eq. (15). With this convention, the paraxial approximation is given by
$k_{\perp,n}(\Omega) \approx \frac{1}{u} [2\omega_0\Omega(1-\frac{u^2}{v_0v_g}) ]^{1/2}$. 

Space-time structured plasma waves (STPs) exhibit properties that are independent of the plasma in which they exist. Unlike conventional plasma waves, which are devoid of correlations in $(\omega,\mathbf{k})$ space and are therefore constrained by the plasma conditions, STPs are constructed with correlations that provide control over their evolution. An example of arbitrary-group-velocity STPs was presented, which was motivated by the subfield of structured light dedicated to controlling the trajectory of peak laser intensity, i.e., spatiotemporal pulse shaping. While much of the analysis from spatiotemporal pulse shaping carries over \cite{Kondakci2019,Palastro2018,Ramsey2023}, unstructured plasma waves are distinct in that their nominal group velocity can be significantly different than their phase velocity. STPs can be realized experimentally, with or without orbital angular momentum, by using the ponderomotive force exerted by two space-time structured laser pulses. More-advanced correlations may allow for STPs with more-exotic structures, such as spatiotemporal optical vortices \cite{Hancock19,Hancock21}. Further work will generalize STPs to magnetized plasma waves, consider STPs driven by charged particle beams, and explore whether STPs can provide control over wave-particle interactions, including linear and nonlinear Landau damping, trapped particle instabilties, or kinetic inflation \cite{Bernstein1957,Kruer1969,Morales1972,Katsouleas1986,Liu1986,Manfredi97,Vu2001,Yin2007,Chapman2012}.

\begin{acknowledgments}

The authors would like to thank A. Raymond, K.L. Nguyen, and T.T. Simpson for discussions.

This report was prepared as an account of work sponsored by an agency of the U.S. Government. Neither the U.S. Government nor any agency thereof, nor any of their employees, makes any warranty, express or implied, or assumes any legal liability or responsibility for the accuracy, completeness, or usefulness of any information, apparatus, product, or process disclosed, or represents that its use would not infringe privately owned rights. Reference herein to any specific commercial product, process, or service by trade name, trademark, manufacturer, or otherwise does not necessarily constitute or imply its endorsement, recommendation, or favoring by the U.S. Government or any agency thereof. The views and opinions of authors expressed herein do not necessarily state or reflect those of the U.S. Government or any agency thereof.

This material is based upon work supported by the Office of Fusion Energy Sciences under Award Numbers DE-SC0023423 and DE-SC00215057, the Department of Energy National Nuclear Security Administration under Award Number DE-NA0003856, the University of Rochester, and the New York State Energy Research and Development Authority.

\end{acknowledgments}

\bibliography{main}% Produces the bibliography via BibTeX.

\end{document}